\documentclass[12pt]{article}

\usepackage{amsmath,amssymb,amsfonts,amsbsy}
\usepackage{cite}
\usepackage{graphicx}
\usepackage{wrapfig}
\usepackage{epsfig}

\usepackage{bbm} 
\usepackage{bm} 
\usepackage{color}                                                       %
\usepackage{dsfont} 
\usepackage{latexsym} 
\usepackage{lscape} 
\usepackage{mathrsfs} 
\usepackage{morefloats} 
\usepackage{floatflt} 
\usepackage{slashed} 
\usepackage{psfrag}

\DeclareFontFamily{OT1}{mygreek}{}%
\DeclareFontShape{OT1}{mygreek}{m}{n}{<->omsegr}{}%
\DeclareFontShape{OT1}{mygreek}{b}{n}{<->omsegrb}{}%
\DeclareFontShape{OT1}{mygreek}{m}{it}{<->omsegri}{}%
\DeclareFontShape{OT1}{mygreek}{bx}{n}{<->sub * mygreek/b/n}{}%
\DeclareFontShape{OT1}{mygreek}{m}{sl}{<->sub * mygreek/m/it}{}%
\DeclareSymbolFont{Greekrm}{OT1}{mygreek}{m}{n}
\DeclareSymbolFont{Greekbf}{OT1}{mygreek}{b}{n}
\DeclareSymbolFont{Greekit}{OT1}{mygreek}{m}{it}
\DeclareMathSymbol{\omegab}{\mathalpha}{Greekbf}{119}
\newcommand{\be}{\begin{equation}}
\newcommand{\ee}{\end{equation}}
\newcommand{\bea}{\begin{eqnarray}}
\newcommand{\eea}{\end{eqnarray}}
\newcommand{\X}{{\vec X}}
\newcommand{\pro}{\partial}
\newcommand{\n}{\hat n}

\newcommand{\hA}{{\hat A}}

\newcommand{\ba}{\begin{array}}
\newcommand{\ea}{\end{array}}

\newcommand{\nn}{\nonumber}
\newcommand{\mn}{{\mu\nu}}

\newcommand{\Int}{\displaystyle\int}

\begin{document}
\addcontentsline{toc}{subsection}{{Consistent gauge invariant nucleon spin decomposition}\\
{\it B.B. Author-Speaker}}

\setcounter{section}{0}
\setcounter{subsection}{0}
\setcounter{equation}{0}
\setcounter{figure}{0}
\setcounter{footnote}{0}
\setcounter{table}{0}

\begin{center}
\textbf{CONSISTENT GAUGE INVARIANT NUCLEON SPIN DECOMPOSITION}

\vspace{5mm}

\underline{D.G. Pak}$^{\,1\,\dag}$ and
P.M. Zhang$^{\,1,2}$

\vspace{5mm}

\begin{small}
  (1) \emph{Institute of Modern Physics, Chinese Academy of Sciences,
 Lanzhou 730000, China} \\
  (2) \emph{Research Center for Hadron and CSR Physics,
 Lanzhou University, Lanzhou 730000, China} \\
  $\dag$ \emph{E-mail: dmipak@gmail.com}
\end{small}
\end{center}

\vspace{0.0mm} 

\begin{abstract}
  We consider a non-uniqueness problem of gauge invariant nucleon spin
decomposition. A gauge invariant decomposition
with a generalized Coulomb constraint for the physical gluon
has been constructed. The decomposition scheme is consistent
with the concept of helicity in non-Abelian gauge theory.
We provide an explicit representation for the gauge invariant
Abelian projection which implies further separation of gluon into
binding and valence parts.
\end{abstract}

\vspace{7.2mm}

It has been a long standing problem of gauge invariant
definition of gluon spin and orbital angular momentum
\cite{pak:jaffe, pak:ji}. Recently a gauge invariant decomposition of
the total nucleon angular momentum into quark and gluon constituents
has been proposed \cite{pak:chen1}, and subsequently other possible
gauge invariant decompositions for nucleon spin
have been suggested \cite{pak:wakam1, pak:cho1}.
Despite on this progress there are still
principal controversies on fundamental conceptual level
in determining a consistent notion for spin
and orbital angular momentum \cite{pak:ji2}.
In the present article we revise the problem of nucleon spin
decomposition and existence of a consistent gauge invariant
concept of spin in the non-Abelian gauge theory.

Let us start with the well known canonical decomposition of
total angular momentum in quantum chromodynamics (QCD)
\bea
&&  J^{can}_{\mu \nu}= \Int d^3 x \big \{\bar \psi \gamma^0 \dfrac{\Sigma_\mn}{2} \psi
-i  {\bar \psi} \gamma^0  x_{[\mu} \pro_{\nu]} \psi
-  \vec A_{[\mu} \cdot \vec F_{\nu] 0}
- \vec F_{0 \alpha} \cdot x_{[\mu} \pro_{\nu]} \vec A_{\alpha} \big \}, \label{canon}
\eea
where we use vector notations for vectors in color space.
All terms in this decomposition, except the first one, are gauge non-invariant.
In the series of papers \cite{pak:chen1} Chen et al have proposed
gauge invariant decomposition of the total angular momentum in
quantum electrodynamics (QED) and QCD.
The basic idea in Chen et al approach is to separate pure gauge and physical
degrees of freedom of the gauge potential in a gauge covariant way
\bea
\vec A_\mu=\vec A_\mu^{pure}+\vec A_\mu^{phys}. \label{split}
\eea
Adding an appropriate surface term
one can obtain the following expression for the total angular momentum tensor
\bea
 J_{\mu\nu}^{can}&=&\int d^3 x \Big \{ \bar \psi \gamma^0 \dfrac{\Sigma_{\mu\nu}}{2} \psi-
i \bar \psi \gamma^0 x_{[\mu} {\cal D}_{\nu]}\psi- \nn \\
&&\vec F_{0[\mu}\cdot \vec A_{\nu ]}^{phys}
-\vec F_{0 \alpha}\cdot x_{[\mu}({\cal D}_{\nu]}
             \vec A_\alpha^{phys}-\vec{\cal F}_{\nu ]\alpha}
             \mbox{\small $(A^{pure})$}) \Big \}, \label{canoncov}
\eea
where ${\cal D}_\mu$ contains a pure gauge field $\vec A_\mu^{pure}$ only.
The given expression for the angular momentum is valid for any split of the gauge potential
before imposing any constraint on physical and pure gauge fields.
In the case of QCD a consistent gauge invariant
decomposition for the nucleon angular momentum
has been proposed by requiring two conditions
on pure gauge $\vec A^{pure}$ and physical $\vec A^{phys}$
components of the gauge field \cite{pak:chen1}
\bea
&&\vec{\cal F}_{\mu\nu} \mbox{\small $(A^{pure})$}=0,~~~~~~~~~~~~~~~~~~~
 {\cal D}_i\vec A_i^{phys}=0, \label{condition2}
\eea
where Latin letters are used for space indices, $i,k,...=(1,2,3)$.
Solving these conditions leads to the gauge invariant decomposition in
the form corresponding to the vector part of (\ref{canoncov}) \cite{pak:chen1}.
In the gauge $\vec A^{pure}=0$ the decomposition
reduces to the canonical one in the Coulomb gauge.
One should notice, that this decomposition is not Lorentz invariant,
so that the notion of gluon spin is frame dependent.
The gauge invariant and Lorentz invariant nucleon spin decomposition
has been suggested in \cite{pak:cho1}. However, in that decomposition the
solving a constraint for the physical gauge potential on mass-shell
encounters a serious problem.

To choose a proper physical nucleon spin decomposition
we require consistence condition
with the helicity notion, which will guarantee the Lorentz invariance.
We will construct explicitly such a
spin decomposition using gauge invariant variables in non-Abelian theory
\cite{pak:perv1}.
The main idea in constructing
gauge invariant variables
is to find a pure gauge $SU(2)$ matrix field in terms
of the initial gauge potential $\vec A$.
Using the equation of motion for the temporal component $A_0^a$
one can write down the equation for the matrix function ${\bf v} \in SU(2)$
\bea
&& \pro_0 {\bf v}(A) = {\bf v}(A) \Big ( \dfrac{1}{D^2(A)} D_j (A) \pro_0 \hA_j -\hat j_0\Big ), \label{eqP1}
\eea
where $\hA_i \equiv A_i^a \tau^a/2$.
Due to the equation of motion for $A_0^a$ it follows that ${\bf v}$ transforms
covariantly, ${\bf v}(A^g)={\bf v} g^{-1}$.
The solution to equation (\ref{eqP1})
can be obtained in the form of time ordered exponent \cite{pak:perv1}
\bea
&& {\bf v}(A)=T \exp \Big {\{} \int^t dt \dfrac{1}{ D^2(A)}  D_j(A) \pro_0 \hA_j-\hat j_0 \Big {\}}.
    \label{Texp}
\eea
This allows to define the gauge invariant variables \cite{pak:perv1}
\bea
\hA_i^I(A)= {\bf v}(A) (\pro_i+\hA_i) {\bf v}^{-1}(A),~~~~
\psi^I(A,\psi)= {\bf v}(A) \psi.    \label{GIvars}
\eea
One can check that $\hA_i^I$ satisfies a constraint which represents a
generalized covariant Coulomb gauge condition
\bea
D_i(A^I) \pro_0 \hA_i^I-\hat j_0 =0. \label{covCG}
\eea
One should stress, that we do not impose this condition,
it follows from the definition of $\hA_i^I$ and $\bf v(A)$.
Finally, from Eqn. (\ref{GIvars}) one finds the following split for the gauge potential
\bea
\hA_i=\hat v^{-1}(A) \pro_i \hat v(A)+ \hat v^{-1}(A) \hA_i^I(A) \hat v(A)
\equiv \hA_i^{pure}+\hA_i^{phys},  \label{splitP}
  \eea
where we can identify the first and second terms as the pure gauge and physical
components needed to make the desired gauge invariant decomposition.
The pure gauge temporal component is defined
by ${\bf v}^{-1}(A) \pro_0 {\bf v}(A)$.
The presented construction of the pure gauge and physical fields in terms
of the unconstrained gauge potential provides an explicit realization
of the gauge invariant nucleon spin decomposition (\ref{canoncov}) with $\vec {\cal F}_{\nu\alpha}(A^{pure})=0$.
The decomposition reduces to the canonical one in the gauge $\vec A^{pure}=0$,
i.e., in the generalized covariant Coulomb gauge (\ref{covCG}).

Let us check consistence of our construction with the concept of helicity.
This will provide frame independent relationship between our gauge invariant definition
of gluon spin density and gluon helicity $\Delta g$ measured in experiment.
The helicity states are described by representations of the little group $E(2)$
\cite{pak:wigner,pak:kim}
which is a subgroup of the Lorentz group. Transformations of the little group
leave the four-momentum of gluon invariant.
If gluon momentum is directed along the $z$ axis, $p_\mu=(\omega,0,0,\omega)$,
the generators of the little group $E(2)$ are given by rotation generator $J_3$
and combinations of Lorentz boost and rotation
\bea
&&N_1=K_1-J_2,~~~~~~~N_2=K_2+J_1.
\eea
The gauge potential represents helicity eigenstates of the operator $J_3$
if the following helicity conditions are satisfied $\vec A_0^{phys}=0, \vec A_3^{phys}=0$ \cite{pak:kim}.
To provide both helicity conditions in a consistent manner with equations of motion
has been an unresolved problem in the case of non-Abelian gauge theory.
In our approach, since one has the first condition $\vec A_0^{phys}=0$
on mass-shell by construction, the second helicity condition
$\vec A_3^{phys}=0$ can be realized by choosing a gauge of either Coulomb or axial or light-cone type.
This is our main result which allows to select a physical gauge covariant
operator $\vec A^{phys}(A)$ and corresponding spin density
consistently with the helicity notion.

Another application of our gauge invariant spin decomposition is
the possibility to provide an explicit representation for the
gauge invariant Cho-Duan Abelian projection \cite{pak:cho3,pak:duan}
which may play an important role in definition of spin decomposition
with dynamic quark momentum \cite{pak:cho1}
\bea
&& \vec{A}_\mu = A_\mu \n +\vec C_\mu + \X_\mu \equiv \hat A_\mu+\vec X_\mu, \nn \\
&& \vec C_\mu = -\dfrac{1}{g} \hat n \times \pro_\mu \hat n, ~~~~~~~~ \vec X_\mu \cdot \hat n =0,
\label{adec}
\eea
where $A_\mu$ is a binding gluon, $\vec X_\mu$ is the valence potential
and $\hat n$ is a unit color triplet. The restricted potential
$\hat A_\mu$ transforms as $SU(2)$ gauge connection,
whereas the valence gluon $\vec X_\mu$ transforms
as a covariant vector.
Let us define the vector $\hat n$ by the relation ${\bf v}(A)=\exp[{i\omega \hat n^i \vec \tau^i}]$.
The pure gauge field $\vec A_\mu^{pure}$ can be constructed in terms of $\hat n$
as follows \cite{pak:plb07}
\bea
&& \vec A_\mu^{pure}=-\tilde C_\mu \n +\vec C_\mu,
\eea
where $\tilde C_\mu$ is a dual magnetic potential.
With this we decompose the gauge potential into pure gauge and physical parts
\bea
&& \vec A_\mu=-\tilde C_\mu \n  +\vec C_\mu + \vec A_\mu^{phys},~~~~~~
 \vec A_\mu^{phys}\equiv \vec A_\mu-\vec A_\mu^{pure} = A_\mu \n +\vec X_\mu.
\eea
Using decompositions (\ref{adec}) one can derive
the expression for the gluon total angular momentum (\ref{canoncov})
which is simplified crucially on mass-shell due to the property $\vec A_0^{phys}=0$,
i.e., $A_0=\vec X_0=0$,
\bea
&&J_{\mu\nu}^{gluon}= \int d^3x \Big \{
-F_{0[\mu } A_{\nu]} -\pro_0 \vec X_{[\mu } \cdot
\vec X_{\nu]}
+F_{0\alpha} x_{[\mu} \pro_{\nu ]} A_\alpha +\pro_0 \vec X_\alpha \cdot
x_{[\mu} \pro_{\nu]}\vec X_\alpha \Big \}, \label{gluonop}
\eea
where $F_{\mu\nu}=\pro_\mu A_\nu-\pro_\nu A_\mu$.
The given decomposition implies further separation of spin densities
of binding and valence gluons. This allows
to define a new spin decomposition with dynamic quark momentum containing
only binding gluon.
By splitting
the gauge potential $\vec A_\mu=\hat A_\mu+\vec X_\mu$
and adding an appropriate surface term to the canonical expression
(\ref{canon}) one results in
a gauge invariant decomposition given by Eqn. (\ref{canoncov})
with replacement
$(\vec A^{pure}_\mu\leftrightarrow \hat A_\mu, ~\vec A^{phys}_\mu\leftrightarrow \vec X_\mu) $.
A new feature of this decomposition is that it does not
contain the spin density for the binding gluon.
Since the binding gluon supposed to
have a dominant gluon contribution to nucleon spin \cite{pak:cho1}, this
supports the experimental data on a small value for
the gluon helicity, $\Delta g \approx 0$.

In conclusion, one has a unique Lorentz invariant
nucleon spin decomposition \cite{pak:cho1},
however, it is not defined on mass-shell and
its  physical meaning remains unclear.
Other known decompositions \cite{pak:chen1, pak:wakam1} are not Lorentz invariant
and lead to frame dependent definitions for gluon spin.
The only gauge invariant and frame independent
notion of spin in the gauge theory is the helicity.
In the present work we have proposed a spin decomposition which is
consistent with helicity concept
and leads to correct expression for the gluon helicity $\Delta g$.

We acknowledge the Organizing Committee for invitation to participate
in the Workshop DSpin-2011. We thank N.I. Kochelev, A.E. Dorokhov, I.V. Anikin, O.V. Teryaev,
A.V. Efremov for numerous discussions and hospitality.
We thank Y.M. Cho, F. Wang and W.M. Sun for critical comments and useful discussions.
The work was supported by NSFC Grants (Nos. 11035006 and 11175215),
and the CAS Visiting Professorship (No. 2011T1J31).

\end{document}